\documentclass[preprint]{iucr}              

\usepackage[pdftex]{graphicx}
\usepackage{amsmath}
\usepackage{listings}

     \journalcode{S}              

\begin{document}                  
\lstset{language=Matlab}


\title{Weighted Least Squares Fit of an Ellipse to Describe Complete or Spotty Diffraction Rings on a Planar 2D Detector}



\cauthor{Michael L}{Hart}{michael.hart@diamond.ac.uk}{address if different from \aff}
\author{Michael}{Drakopoulos}

\aff{Diamond Light Source Ltd, Diamond House, Harwell Science and Innovation Campus, Didcot,
Oxfordshire OX11 0DE. \country{United Kingdom}}









\maketitle                        

\begin{synopsis}
Weighted least squares algebraic fit of complete or spotty X-ray powder diffraction rings.
\end{synopsis}

\begin{abstract}
We present a procedure for fitting an ellipse to powder diffraction patterns recorded on a planar 2D detector, which uses the peak intensities as weights. This procedure works for complete and spotty diffraction rings. We outline all the steps required: the interpolation of Cartesian pixel data into polar co-ordinates based upon an approximate ellipse centre; the identification of peak locations along each radial direction of a diffraction ring; and the weighted least squares algebraic fit of an ellipse to that diffraction ring. The performance of the procedure is assessed by characterising a complete diffraction ring from a powder standard, and a spotty ring generated from a complete ring of the standard. The fit of an ellipse to a spotty ring from an alkane complex is also shown. Agreement in the calculated parameters for the two rings is good. This procedure is implemented as part of an X-ray beam energy and 2D detector calibration routine which is currently in use at Beamline I12 (Diamond Light Source, UK).
\end{abstract}


\section{Introduction}

The analysis of powder diffraction patterns as ellipses can yield information on a sample's strain \cite{bronfenbrenner11,hanan04}, and provide a means for calibrating the image plane in relation to the sample \cite{hart13b,hong12,ilavsky12,gommes10,cervellino08,hammersley96}. Any powder diffraction ring incident on a planar 2D detector is a conic section, and in the case of an off orthogonal detector, is an ellipse. The use of such patterns allows for the correlation problem between the X-ray beam energy and the sample-to-detector distance to be broken, and for all instrument parameters to be calculated and solved, \cite{hart13b}. 

Diffraction from a powder sample produces a diffraction cone with the sample at its apex. If the sample is strain free then the cone will be right circular, but if the sample exhibits a first order strain then the cone will be right elliptical. In general a planar 2D detector placed to be `orthogonal' to the incoming X-ray beam will still exhibit a small degree of tilt \cite{hart13b,cervellino08} as perfect orthogonality is difficult to achieve, however in both cases - unstrained and strained - the intersection between a cone and a tilted plane is an ellipse, \cite{eberly12}. As such a diffraction ring recorded on a planar 2D detector can always be described as an ellipse.

We present a detailed technique for fitting an ellipse to complete or spotty rings. We choose to weight the fit of an ellipse based upon the intensity of the data point, although we note that other choices for the weights may used, such as the full width at half maximum. By demonstrating the fit of an ellipse to both simulated and spotty rings, we do not suggesting that the fit of an ellipse is appropriate for use with all spotty patterns, but leave it to the reader to decide if the information obtained from the fit of an ellipse is representative of the underlying data. We note however that this technique is sufficiently accurate for use on complete diffraction rings, as demonstrated elsewhere, \cite{hart13b}.

Within this presentation we frequently reference another piece of work authored by ourselves, this is because the other piece of work was developed in conjunction with that presented here. 

\section{Ellipse Fitting Routine}

Our method, to fit a weighted ellipse to a diffraction ring as recorded on a planar 2D detector (the image), involves several steps: an approximate centre to the rings is found from the image; using this centre as an origin, the intensity data is interpolated into polar coordinates with equal intervals in azimuth; peak locations are determined by fitting Gaussians to the data in polar coordinates; and finally the Cartesian coordinates of the peak centres are fit with an ellipse using the fitted peak intensities as weights.

\subsection{Finding an Approximate Centre}

The projected location of the incident beam ($Q=0$ \AA$^{-1}$) does not coincide with either the ellipse centre or the focal point of the ellipse, \cite{hart13b}. However, our method only requires an image co-ordinate that is approximately central to the diffraction ring that is to be described. The same approximate centre can be used for all diffraction rings present in an image for small detector tilts. 

We use the local maxima of the self-convolution of the diffraction image to calculate a co-ordinate for the approximate centre \cite{cervellino08}, with suitable downsampling to help reduce computation time. This method is both mathematically elegant and robust, although requires that at least half of each diffraction ring be recorded on the detector in order to work, and some degree of symmetry if the rings are spotty.

\subsection{Interpolation of Cartesian Data along an Arbitrary Line}

Having found an approximate centre which can be used as an origin, $O$, the intensity data which is referenced in the Cartesian coordinates, $(x,y)$, of the detector is transformed  into polar coordinates, $(r,\phi)$, using bilinear interpolation. In polar coordinates, we consider $R$ rays originating from $O$, equally spaced in azimuth, $\phi$. The interval along the radial coordinate is set to be the same length as the width of an individual detector pixel (in the case of the width of a diffraction ring spanning several pixels). $R=720$ is sufficient for complete rings, but more rays may be required if the rings are spotty (e.g. $R=3000$). When rings are spotty it is important to have enough rays such that all spots from a diffraction ring intersect a ray, and are thus statistically included.

\subsection{Approximate Identification of Peak Locations}

It is often the case that multiple diffraction peaks are present within the image, and thus multiple peaks may be present along each of the $R$ rays. In the case of complete rings there will be a one-to-one correspondence between the peaks in the $R$ rays and the peaks in the image. In the case of spotty rings, different peaks will be absent from the $R$ rays. Before a peak can be assigned a location in $r$ and $\phi$, it is necessary to isolate the intensity data around that peak. To do this we define annuli such that the each diffraction ring lies completely within its own annulus. 

Along each of the $R$ rays, the $r$ coordinate of the $N$ most intense local maxima are determined. A histogram of the local maxima locations across all $R$ rays is then created, giving a plot of peak frequency versus $r$. Approximate peak locations can then be identified from the most intense local maxima in the histogram. In the case of spotty rings where the spots from different diffraction rings tend not to appear at the same azimuth, $N$ can be chosen to be less than the number of rings in the image, this reduces the potential of false positives being identified along each of the $R$ rays. Finally annuli can be defined based upon the approximate peak locations.

This method works well on complete or spotty rings, and is designed to work well if the data is noisy and/or of low contrast, and in cases where neighbouring rings may be poorly separated. A difficulty can however arise when two or more rings of high eccentricity (perhaps due to a large detector tilt) are close together, such that the semi-major axis of the smaller ellipse is similar is size or greater than the semi-minor axis of the larger ellipse. This problem can be overcome by using elliptical rather than circular annuli. The switch to elliptical annuli, in real space, can be achieved by defining contours of constant $Q$, which can be generated using approximate instrument calibration parameters. 

\subsection{Accurate Identification of Peak Locations}

The previous section established annuli to define regions in intensity data which each contain a single peak. Using the original raw data, we now determine peak locations along each of the $R$ rays.

All peaks along all rays can then be assigned an accurate location in $r$ and $\phi$ by fitting a Gaussian to each region of intensity data. We define the accurate location of the peak in $r$ as the centre of the fitted Gaussian, and also define the weight of the peak as the maximal value of the fitted Gaussian\footnote{Alternative functions can be used to describe the diffraction peaks. In the case of two or more peaks being close enough together such that independent annuli cannot be assigned, it may necessary to employ an algorithm for extracting the locations of individual reflections.}. For the purpose of fitting ellipses to the diffraction rings, the locations of all peaks are converted into the Cartesian coordinates of the detector, $(x,y)$.

For spotty rings it can be of benefit to implement a procedure for discarding certain subsets of peak locations. As spotty rings are incomplete, some regions of intensity data will be representative of noise. We discard peak locations which derive from fitted Gaussians whose maxima lie outside of the local annulus, or if the underlying data is of low intensity. By design the intensity weighted ellipse fitting routine described in the next section is somewhat robust to spurious data points by virtue of such data points having a low intensity or weight, however even low intensity spurious data points will influence a fit, and thus should be discarded.

\subsection{Fitting an Intensity Weighted Ellipse to a Diffraction Ring}

\subsubsection{Theory}

An ellipse is fitted to the Cartesian coordinates of the accurately determined peak locations. The fit is weighted by the previously calculated peak intensities. In Cartesian coordinates, the equation for an ellipse can be given by an implicit second order polynomial:

\begin{equation}\label{equ:ellipse}
F\left(\textbf{a},\textbf{x},f\right) = \textbf{a}.\textbf{x} + f = ax^2 + bxy + cy^2 + dx + ey + f = 0
\end{equation}

where $\textbf{a} = \left[a\ b\ c\ d\ e\right]^T$ and $\textbf{x} = \left[x^2\ xy\ y^2\ x\ y\right]^T$.

The distance from peak location $\left(x_{i},y_{i}\right)$ to the conic, $F\left(\hat{\textbf{a}},\textbf{x},f\right)=0$, is the `algebraic distance', \cite{fitzgibbon99}. To fit an ellipse where each of $N$ peak locations is associated with a weight, $w_i$, we choose to find the estimated parameter values of the best fit weighted ellipse by minimising the weighted sum of squared algebraic distances of the curve to the $N$ peak locations $\textbf{x}_{i}$ \cite{haralick92},

\begin{equation}\label{equ:minim}
\hat{\textbf{a}} = \underset{\textbf{a}}{\arg\min}\sum^{N}_{i=1}{w_{i}F(\textbf{x}_i)^2} = \underset{\textbf{a}}{\arg\min}\sum^{N}_{i=1}{w_{i}\left|f_{i}+\sum^{5}_{j=1}{X_{ij}a_{j}}\right|^2}
\end{equation}

where $X_{ij}$ is the appropriate element of the matrix $\textbf{X}$ given by, $\textbf{X} = \left[\textbf{x}_1\ \textbf{x}_2\ \ldots\ \textbf{x}_N\right]^T$. To avoid the trivial solution of $\hat{\textbf{a}} = 0_5$ and $f=0$ we use the constraint $f=constant$, as investigated by elsewhere, \cite{rosin93}. Solving equation (\ref{equ:minim}) gives an algebraic expression for the best fit estimated parameter values, 

\begin{equation}\label{equ:solution}
\hat{\textbf{a}} = (\textbf{X}^{T}\textbf{W}\textbf{X})^{-1}\textbf{X}^{T}\textbf{W}
\end{equation}

where $\textbf{W}$ is a diagonal matrix of the $N$ weights associated with each peak location $\textbf{x}_i$. We have used the constraint $f=-1$. 

\subsubsection{Implementation within Matlab}

The unweighted solution for fitting an ellipse has been implemented within a Matlab (MathWorks) function, \cite{gal03}. This excellent implementation uses the constraint $f=constant$, and a pre- and post-fit transformation to avoid data passing through the origin and thus reduces numerical errors and ensures invariance to translations of the data. The function gives parameters for the ellipse in terms of the ellipse centre, the tilt direction and the lengths of the semi-minor and semi-major axes. 

A direct implementation of equation (\ref{equ:solution}) can be computationally slow, and so we perform an equivalent calculation by modifying Ohad Gal's function with the following code changes,\newline

\begin{lstlisting}
X = [x.^2, x.*y, y.^2, x, y];
a = sum(X)/(X'*X);
\end{lstlisting}
to \newline
\begin{lstlisting}
X = [x.^2, x.*y, y.^2, x, y];
V = w*ones(1,5);
X2 = V.*X;
a = sum(X2)/(X'*X2);
\end{lstlisting}

where w is an additional input to the function, and is a column vector containing the weights $w_{i}$.


\section{Results and Discussion}

\begin{figure}\label{fig:fit}
\caption{Using a cerium dioxide standard, peak locations (red dots) and an arc of the fitted ellipses (red curve) are shown for a complete (a,c) and corresponding simulated spotty ring (b,d).}
\includegraphics[width=12cm]{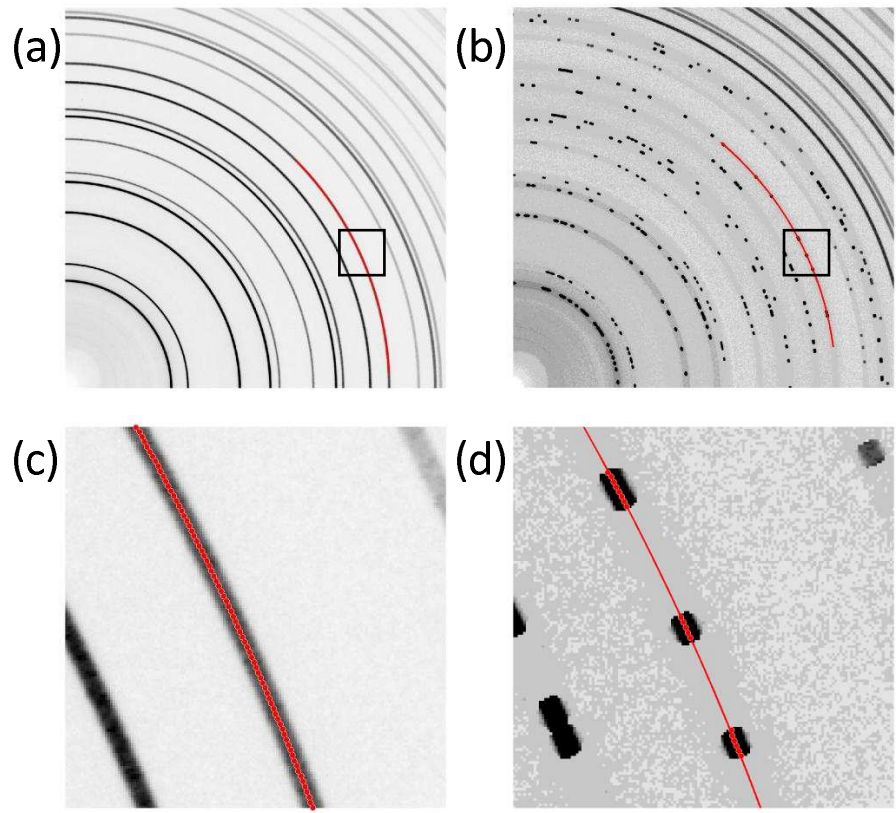}
\end{figure}

Figure \ref{fig:fit} demonstrates the quality of fit for complete and spotty diffraction rings. Figure \ref{fig:fit}a shows a quarter of the powder diffraction pattern from a cerium dioxide standard (NIST - Standard Reference Material 674b) captured on a Pixium (2880 by 2881 pixels) flat panel detector (RF4343, Thales), with a pixel size of $148 \mu m$ by $148 \mu m$. The diffraction pattern was recorded at the Joint Engineering, Environmental and processing beamline (I12) at Diamond Light Source Limited (UK). An arc of the fitted ellipse is shown for one of the powder diffraction rings. For comparison purposes we generate a spotty diffraction pattern from the complete diffraction pattern. This ensures that the complete and spotty diffraction rings perfectly overlap and are thus suitable for comparative purposes. We generate the spotty pattern (rings $1$ to $13$) by modulating the intensity of the pixels in a complete ring with $100$ randomly located Gaussian envelopes. This modulated spotty pattern is shown in figure \ref{fig:fit}b, together with an arc of a fitted ellipse. Figures \ref{fig:fit}c and \ref{fig:fit}d show the fine detail of these fits, where the red dots show the peak locations and the red curves show the fit of the intensity weighted ellipse to those peak locations. In the case of the spotty pattern it can be seen that peak locations between the spots have been dropped due to a lack of information. At the edges of the spots it can be seen that the spots can deviate in position from the true peak location, however due to these data points being associated with a low intensity, and thus a low weight, these peak locations will not significantly contribute to the fitted ellipse.

The parameters of the fitted ellipses, for the 10th ring of the complete and spotty patterns, compare well: in pixels, the fitted values for the semi major axis are $1014.05$ and $1014.06$ respectively; and the co-ordinates of the ellipse centres are $(1434.21,1441.93)$ and $(1434.27,1441.95)$. Positional agreement is at the $10$ micrometre level. The 10th spotty ring has $17\%$ of the total intensity of the 10th complete ring.

\begin{figure}\label{fig:spotty}
\caption{Figures (a,b,c, and d) demonstrate the fitting of an ellipse to a single ring from an alkane complex 2D spotty diffraction pattern, \cite{pike12}. Zoomed in regions taken from each quadrant of the spotty ring overlaid with the peak locations (red dots) and fitted ellipse (red curve) are shown.}
\includegraphics[width=12cm]{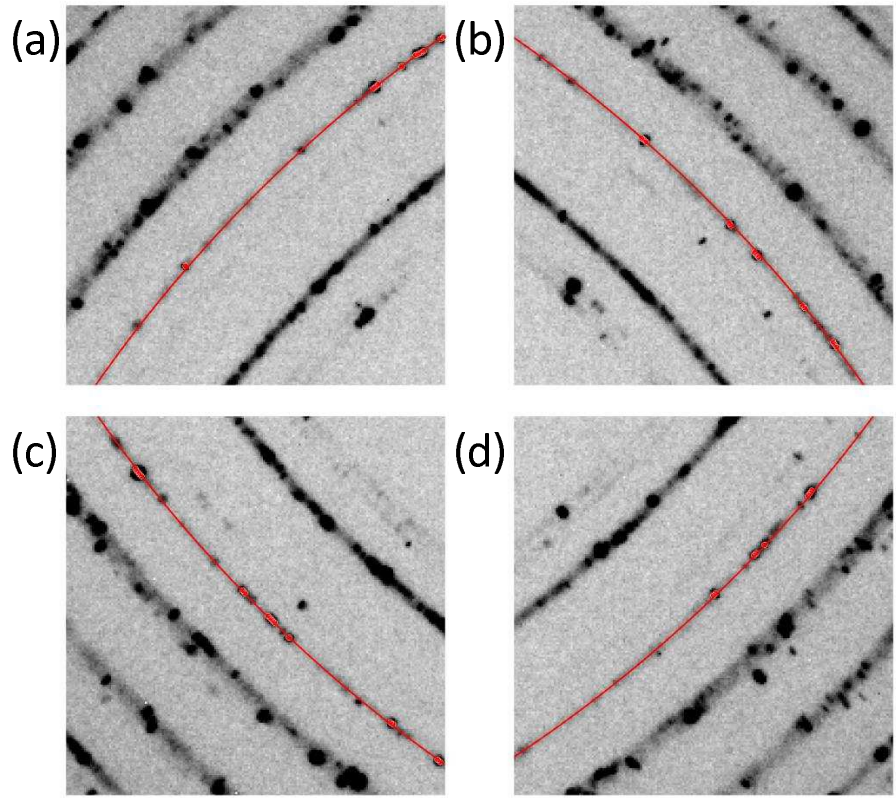}
\end{figure}

Figure \ref{fig:spotty} shows (X-ray energy, $52.14 keV$; sample to detector distance, $2510.1 mm$) the fit of an ellipse to a spotty diffraction ring from an NBD (norbornadiene) complex: [Rh(NBD)(iPr2CH2CH2iPr2)][BArF4] (ArF = 3,5-(CF3)2C6H3), \cite{pike12}. The spots present in this ring do not possess a single value of $Q$, however the fitted ellipse will be representative of the spots present.

An alternative method of directly fitting a single weighted ellipse using the pixel intensities in the Cartesian coordinate system of the flat panel detector, within an elliptical annulus, was also considered. Such an approach would be more in line with a method described elsewhere \cite{hanan04}. This alternative approach fits several unweighted ellipses to diffraction ring pixel data segregated according to intensity band, and then determines a final ellipse to describe the ring as an intensity weighted average of all the unweighted ellipses. We believe that such an approach would be problematic in the case where two diffraction rings had little separation in $2\theta$. The elevated pixel intensity between two poorly separated Debye-Scherrer rings will cause either fitted ellipse to be skewed towards the midpoint of the two rings. Our approach of fitting to the peaks of the each diffraction ring avoids this potential problem.

Our method for describing a diffraction ring as an ellipse is currently applied as part of a calibration routine to find the energy and instrument parameters for a 2D detector, \cite{hart13b}, as used at beamline I12 (Diamond Light Source, UK). This geometrical and analytical routine calculates the energy, the sample-to-detector distance, and the detector's orientation and position - all the parameters required to perform a 2D to 1D data reduction. The accuracy of our ellipse fitting method helps the calibration routine to resolve peak location at the $\Delta Q / Q\approx10^{-5}$ level.

\section{Conclusion}

We have a demonstrated a technique for describing a complete or spotty diffraction ring as an ellipse. By introducing weights to the fit of the ellipse, this technique becomes robust enough to describe spotty diffraction rings with minimal loss in accuracy. With help from Ohad Gal's ellipse fitting script, \cite{gal03}, we show how a single weighted ellipse can be algebraically fit to a diffraction ring.






\referencelist[refs]



\newpage

\end{document}